\documentclass[10pt,aps,pre,onecolumn,superscriptaddress,showpacs,preprintnumbers,amsmath,amssymb,notitlepage]{revtex4-1}

\usepackage{amsmath}
\usepackage{graphicx}
\usepackage{dcolumn}
\usepackage{bm}
\usepackage{color}

\begin{document}

\title{Codon Bias Patterns of {\it E.coli}'s Interacting Proteins} 

\author{Maddalena Dilucca}
\email{maddalena.dilucca@roma1.infn.it}
\affiliation{Dipartimento di Fisica, Universit\`a ``Sapienza'', Rome, Italy}
\author{Giulio Cimini}
\affiliation{Istituto dei Sistemi Complessi (ISC)-CNR, UoS Universit\`a ``Sapienza'', Rome, Italy}
\author{Andrea Semmoloni}
\affiliation{Dipartimento di Fisica, Universit\`a ``Sapienza'', Rome, Italy}
\author{Antonio Deiana}
\affiliation{Dipartimento di Fisica, Universit\`a ``Sapienza'', Rome, Italy}
\author{Andrea Giansanti}
\affiliation{Dipartimento di Fisica, Universit\`a ``Sapienza'', Rome, Italy}
\affiliation{INFN Roma1 unit, Rome, Italy}
\date{\today}

\begin{abstract}
Synonymous codons, {\em i.e.}, DNA nucleotide triplets coding for the same amino acid, are used differently across the variety of living organisms. 
The biological meaning of this phenomenon, known as {\it codon usage bias}, is still controversial. In order to shed light on this point, we propose a new codon bias index, $CompAI$, 
that is based on the competition between cognate and near-cognate tRNAs during translation, without being tuned to the usage bias of highly expressed genes. 
We perform a genome-wide evaluation of codon bias for {\it E.coli}, comparing $CompAI$ with other widely used indices: $tAI$, $CAI$, and $Nc$. 
We show that $CompAI$ and $tAI$ capture similar information by being positively correlated with gene conservation, measured by ERI, and essentiality, whereas, 
$CAI$ and $Nc$ appear to be less sensitive to evolutionary-functional parameters. Notably, the rate of variation of $tAI$ and $CompAI$ with ERI allows to obtain sets of genes 
that consistently belong to specific clusters of orthologous genes (COGs). We also investigate the correlation of codon bias at the genomic level 
with the network features of protein-protein interactions in {\it E.coli}. We find that the most densely connected communities of the network share a similar level of codon bias (as measured by $CompAI$ and $tAI$). 
Conversely, a small difference in codon bias between two genes is, statistically, a prerequisite for the corresponding proteins to interact.
Importantly, among all codon bias indices, $CompAI$ turns out to have the most coherent distribution over the communities of the interactome, 
pointing to the significance of competition among cognate and near-cognate tRNAs for explaining codon usage adaptation.
\end{abstract}

\maketitle 

\section*{Introduction}

The genetic information carried by the mRNA and then translated into proteins is encoded into nucleotide triplets called \emph{codons}. 
Four alternate nucleotidic bases (A,U,C,G) compose the mRNA, so that there are $4^3=64$ possible codons that have to code for only $20$ naturally occurring amino acids. 
The genetic code is therefore redundant: while a few amino acids correspond to a single codon, most amino acids can be encoded by different codons. 
Different codons coding for the same amino acid are known as synonymous codons, and in a wide variety of organisms synonymous codons are used with different frequencies---a phenomenon known as \emph{codon bias}. 
With the advent of whole-genome sequencing of numerous species, genome-wide patterns of codon bias are emerging in the different organisms. Various factors such as 
expression level, GC content, recombination rates, RNA stability, codon position, gene length, environmental stress and population size, can influence codon usage bias within and among species \cite{Hershberg2008}. 
While the biological meaning and origin of codon bias is not yet fully understood, there is a large consensus that the degeneration of the genetic code might provide an additional degree of freedom 
to modulate accuracy and efficiency of translation \cite{Shabalina2013}. Indeed, population genetic studies \cite{Tuller2014} have shown that synonymous sites are under weak selection, 
and that codon bias is maintained by a balance between mutation-selection (random variability in genetic sequences followed by fixation of the optimal codons) and genetic drift 
(allowing for the occurrence of non-optimal codons). In fact, highly expressed genes feature an extreme bias by using a small 
subset of codons, optimized by translational selection \cite{Bennetzen1982,Gouy1982,Jansen2003}. On the other hand, the persistence of non-optimal codons in less-expressed sequences causes long breaks 
during protein synthesis; this could be the result of genetic drift and have a key role in the protein folding process \cite {Weili2014, Pop2014}. In addition, codon usage appears to be structured along the genome, 
with neighboring genes having similar codon compositions \cite{Daubin2003}, and codon bias seems positively correlated to gene length 
(as a result of selection for accuracy in the costly production of long proteins) \cite{Eyre-Walker1996}. 
In the last years there has been a wide effort in developing effective ways to measure codon bias \cite{Roth2012}. The most widely used indices include 
the {\em Codon Adaptation Index} ($CAI$) \cite{Sharp1987}, the {\em tRNA Adaptation Index} ($tAI$) \cite{dosReis2004}, and the {\em Effective Number of Codons} ($Nc$) \cite{Wright1990}, 
each of them having specific advantages and drawbacks. For instance, $CAI$ and $tAI$ correlate well with gene expression levels, however such correlation is a natural consequence of their definition: 
they are tuned on a reference set of highly expressed genes. $Nc$ is instead basically a measure of the entropy of the codon usage distribution, and thus shows a lower correlation with expression levels. 

In this work we propose a novel codon bias index named {\em Competition Adaptation Index} ($CompAI$), which does not rely on information about gene expression levels, but instead has a self-consistent 
biological meaning---based on tRNA availability and competition between cognate and near-cognate tRNAs. In other words, $CompAI$ is a parameter-free index that does not require a set of reference genes 
for its calibration, a fact that constitutes its main advantage with respect to $CAI$ and $tAI$. Moreover, $CompAI$ is designed to extract genetic signals that could be directly correlated 
to experimental measures for translation speeds, an emerging and challenging issue still to be explored. 
In order to show the advantage of the novel codon bias index, we perform a genome-wide comparison of $CAI$, $tAI$, $Nc$ and $CompAI$ for {\em Escherichia Coli} (E.coli). 
Our analysis reveals that the information on gene conservation across species and gene essentiality is better captured by codon bias metrics that build on tRNA availability ($tAI$ and $CompAI$). 
We also study codon bias in relation to the connectivity patterns of the protein-protein interaction network (PIN) \cite{Szklarczyk2011} of E.coli. 
We thus show that translational selection systematically favors proteins with the highest number of interactions and belonging to the most densely connected community of the network, 
at least when the bias is measured by $CompAI$ and, to a smaller extent, by $tAI$. Additionally, we address the issue of how much a similarity in the codon usage bias of a set of genes 
is reflected on the interactions among the corresponding proteins. A principal component analysis for the variability of codon bias indices indeed reveals that 
closeness of a set of genes in the space of the two principal components likely results in the corresponding proteins to interact---in comparison with an appropriate null model. 

Overall, our study reveals that $CompAI$ captures more information than the other indices about the connection between codon bias and the topology of the interactome. 
Besides, we recall that $CompAI$ does not require calibration on gene expression levels and has a consistent biological meaning based on the competition between cognate and near-cognate tRNAs.
These observations stress the potential of the new index to both measure and explain codon usage bias.

\section*{Materials and Methods}

\subsection*{Sequences}

In this work we investigate the genome of E.coli K-12 substr. MG1655, whose 4005 coding mRNA sequences have been collected from GenBank \cite{Benson2012}. 
The gene copy numbers coding for each tRNA (tGCN) were derived from the Genomic tRNA database \cite{Chan2009}.

\subsection*{Conservation and Essentiality of E.coli genes}

In order to have an index for gene ``conservation'', we use the normalized Evolutionary Retention Index (ERI) \cite{Gerdes2003}: for each gene in E.coli's genome, its ERI measures how much that gene is shared 
among other 32 bacterial species (having at least an ortholog of the given gene). A low ERI value thus denotes that a gene is specific to E.coli, whereas, high ERI is characteristic 
of highly shared (and therefore conserved) genes. 
Concerning gene ``essentiality'', we use the classification of Gerdes et al. \cite{Gerdes2003} for the E.coli genome into 606 essential and 2940 non-essential genes, 
based on experimental measures of gene resistance against transposon insertion.

\subsection*{Codon Bias Indices}

Codon usage bias can be assessed, for each gene in a given genome, by various indices that can be classified into broad groups based on: 
(i) codon frequencies; (ii) reference gene sets; (iii) deviation from a postulated distribution; (iv) information theory; (v) interactions among tRNAs (see \cite{Roth2012} for an overview). 
We focus here on the most widely used indices: $tAI$ \cite{dosReis2004}, that belongs to groups (ii) and (v) by requiring calibration on a set of highly expressed genes; $CAI$ \cite{Sharp1987}, 
a group (i) and (ii) index built on local statistics of codon usage and on a reference list of optimally expressed genes; $Nc$ \cite{Wright1990}, a group (i) index based on the number of different codons 
used in a coding sequence. The novel codon bias index we propose in this work, $CompAI$, is instead based on the competition of cognate and near-cognate tRNAs to bind to the A-site on the ribosome 
during translation, and is thus a group (v) index that does not need tuning on a reference set of highly expressed genes. While the formal definition for $CompAI$ and the rationale behind are given below, 
we refer to the Supporting Information for the operative definition of $CAI$, $tAI$ and $Nc$. 

{\bf Competition Adaptation Index} ($CompAI$). It is generally accepted that translation speed depends on the efficiency of the codon/anticodon pairing in the A site of the ribosome \cite{Rodnina2001}. 
Hence, for a given codon, the rate of amino acid synthesis is essentially influenced by two dominant processes: the number of collisions of the corresponding tRNA with the ribosomal A site 
(which strongly depends on tRNA concentration in the cell) and the specificity of the codon-anticodon pairing. Such a pairing process satisfies the Watson-Crick base-pairing rules (G-C and A-U, and vice versa) 
for the first two bases, whereas, the rule on the third (or {\em wobble}) base is more relaxed and non-standard pairing is allowed in some cases (wobble complementarity)\cite{Crick1966}. 
Hence, there are cases in which several tRNAs pair with the same codon (provided that these are identical in the first two bases) and are called isoacceptor or cognate tRNAs. 
Codon-anticodon interactions are thus characterized by competition between cognate tRNA (with perfect or wobble complementarity between mRNA codon and tRNA anticodon), 
near-cognate tRNA (with a mismatch in only one of the first two bases) and non-cognate tRNA (with at least two mismatches). 
Discrimination between correct and wrong tRNA according to base pairing features very high fidelity 
(error rate $f\sim10^{-3}\div 10^{-4}$). Rejection of the wrong tRNA can occur in two distinct phases \cite{Rodnina2001,Gromadski2004}: 
initial selection of the ternary complex EF-Tu-GTP-aa-tRNA and subsequent proofreading of aa-tRNA after GTP hydrolyzation. 
The first interaction is fast and does not depend on the choice of codons, in order to allow the ribosome to quickly screen for the available tRNAs. 
The second step is instead sensitive to base complementarity, featuring the first selection between cognate and near-cognate tRNAs: 
non-cognate are excluded almost immediately with $f\sim 10^{-1}$, and then a more strict and efficient proofreading takes place, excluding near-cognate with $f\sim 10^{-2}$. 
This means that near-cognate tRNA (unlike non-cognate tRNA) can enter into the interaction process between the ternary complex and the site A of the ribosome, 
and (when not accepted, in very few cases) can be rejected at the stage of initial recognition or during proofreading. In any event, this process results in a time delay of translation, 
because near-cognate rejection brings the ribosome back to the initial state of waiting for the correct tRNA.

The rationale behind the definition of $CompAI$ is precisely that of building an index which is based both on tRNA availability and on competition between cognate and near-cognate tRNAs 
that could modulate the speed of translation of mRNAs into proteins. Note that, since in vivo experimental determinations of tRNA concentrations 
are available only for few organisms, we will implement $CompAI$ using the number of tRNA gene copies (tGCN) which, at least in simple organisms, 
has a high and positive correlation with tRNA abundance \cite{Ikemura1981,Dong1996,Kanaya1999,Percudani1997} (a similar approach is adopted in the definition of $tAI$ \cite{dosReis2004}). 
For each codon $i$ we define its absolute adaptiveness value ($W_i$) as:
\begin{equation}\label{eq:ci}
W_i=\left(\overset{m_i}{\underset{j=1}{\sum}}\mbox{tGCN}_{ij}\right)\left[\frac{\overset{m_i}{\underset{j=1}{\sum}}\mbox{tGCN}_{ij}}{\overset{m_i}{\underset{j=1}{\sum}}\mbox{tGCN}_{ij}+\overset{m_i^{nc}}{\underset{j=1}{\sum}}\mbox{tGCN}_{ij}^{nc}}\right].
\end{equation}
Here $m_i$ is the number of isoacceptor tRNA sequences (anticodons) that recognize codon $i$ ({\em i.e.}, containing either the anti-codon $i$ or all its cognates that are read by $i$)
and $\mbox{tGCN}_{ij}$ is the gene copy number of the $j$-th of such tRNAs, whereas $m_i^{nc}$ is the number of tRNA sequences that are near-cognate of $i$ 
and $\mbox{tGCN}_{ij}^{nc}$ is the gene copy number of the $j$-th of such tRNAs (see also Fig. \ref{fig.cognate} of Supporting Figures). 
The amount in square brackets represents a penalty introduced by the competition with near-cognate tRNAs, assuming unit or zero values in the cases of smaller and higher competition, respectively. 
This term thus assumes the role of selective constraint on the efficiency of the codon-anticodon coupling. 
Importantly, and at odds with $tAI$, these terms do not result from optimization on expression levels, but have a biological justification based on cognate/near-cognate competition. 
Note that, in the computation of $W_i$ for a given codon, we count as isoacceptor tRNAs those with perfect or wobble base pairing that also carry the same amino acid of $i$'s anticodon. 
Computation of $CompAI$ continues by defining for each codon $i$ its relative adaptiveness value $w_i=W_i/W_{max}$, 
where $W_{max}$ is the maximum value between all the $W_i$ of codons.
$CompAI$ of gene $g$ is finally defined as the harmonic mean of the relative adaptiveness of its codons:
\begin{equation}
CompAI_g=\frac{l_g}{\overset{l_g}{\underset{i=1}{\sum}}w_i^{-1}} \label{eq:compAI}
\end{equation}
The choice of the harmonic mean (rather than geometric as for $CAI$ and $tAI$) is consistent with the association of $CompAI$ with the rate of protein synthesis. 
Indeed, the translation speed of codon $i$ can be defined as the reciprocal of the concentration of the corresponding tRNA isoacceptors \cite{Zhang2009}. 
Therefore, if codon $i$ is read at a speed proportional to $w_i$, then the average translation speed of a gene is given by the harmonic mean of the $\{w_i\}$ associated to its codons. 
$CompAI$ takes values between 0 and 1, where values close to 0 (1) indicate highest (lowest) competition, and therefore a low (high) translation rate. 

\subsection*{Protein-Protein Network Analysis}

In this study we use protein interaction data collected in STRING (Known and Predicted Protein-Protein Interactions) \cite{Szklarczyk2011}. 
In such database, each predicted interaction is assigned with a confidence level or probability $w$, evaluated by comparing predictions obtained by the different techniques 
\cite{Chien1991,Puig2001,Phizicky1995} with a set of reference associations, namely the functional groups of KEGG (Kyoto Encyclopedia of Genes and Genomes) \cite{Kanehisa2000}. 
In this way, interactions with high $w$ are likely to be true positives, whereas, a low $w$ likely corresponds to a false positive. 
Since the percentage of false positives can be very high \cite{Huang2007}, we select a stringent cut-off $\Theta=0.9$ that allows a fair balance between coverage and interaction reliability 
(see the probability distribution $P(w)$ in the left panel of Fig. \ref{fig.P_kw} of Supporting Figures). 
We thus build the protein-protein interaction network (PIN) of E.coli by placing a link between each pair of proteins (nodes) $i,j$ provided that $w_{ij}>\Theta$. 
The resulting number of connections or {\em degree} for a given protein $i$ is denoted as $k_i$. 

To detect communities of PIN we resort to Molecular Complex Detection (MCODE) \cite{Bader2003}. In a nutshell, MCODE iteratively groups together neighboring nodes 
with similar values of the core-clustering coefficient, which for each node is defined as the density of the highest $k$-core of its immediate neighborhood times $k$ \cite{foot1}.
MCODE detects the densest regions of the network and assigns to each found community a score that is its internal link density times the number of nodes belonging to it \cite{foot2}. 
We also characterize each found community $c$ with the mean value $\bar{x}_c$ and standard deviation $\sigma_c$ 
of codon bias values within the community, and use them to compute a $Z$-score as $Z_c=(\bar{x}_c-\bar{x}_n)/\sqrt{\sigma_c^2+\sigma_n^2}$ (where $\bar{x}_n$ and $\sigma_n$ are, respectively, 
the mean value and standard deviation of codon bias values computed on the whole network). In this way, a value of $Z_c>1$ ($Z_c<-1$) indicates that community $c$ features significantly higher (lower) 
codon bias than the population mean.

Finally note that each node of PIN is photogenically classified according to the Clusters of Orthologous Groups \cite{foot3} (COGs) of proteins \cite{Tatusov2001}. 
COGs are generated by comparing predicted and known protein sequences in all completely sequenced genomes to infer sets of orthologs. 
Each COG consists of a group of proteins found to be orthologous across at least three lineages and likely corresponds to an ancient conserved domain \cite{Tatusov2001}.

\subsection*{Principal Component Analysis}

Principal Component Analysis (PCA) \cite{Jolliffe2002} is a multivariate statistical method to transform a set of observations of possibly correlated variables into a set 
of linearly uncorrelated variables (called principal components) spanning a space of lower dimensionality. 
The transformation is defined so that the first principal component accounts for the largest possible variance of the data, and each succeeding component in turn has the highest variance possible 
under the constraint that it is orthogonal to ({\em i.e.}, uncorrelated with) the preceding components. 

We use this technique on the space of codon bias indices, so that each gene of E.coli is represented as a 4-dimensional vector with coordinates ($CompAI$, $CAI$, $tAI$, $Nc$). 
Such coordinates are separately normalized to zero mean and unit variance over the whole genome. We then obtain the associated covariance matrix between the four dimensions of codon bias and diagonalize it. 
The eigenvectors of the covariance matrix, ordered according to the magnitude of the corresponding eigenvalues, are the principal components of the original data. 

\subsection*{Configuration Model}

In order to assess how significant are the codon usage patterns observed for the PIN, we need to compare the E.coli interactome with a suitable null model for it, 
{\em i.e.}, an appropriate randomization of the network. Here we follow the most common approach in statistical mechanics of networks of using the {\em Configuration Model} (CM) \cite{Park2004}. 
The basic idea is to build the null model as an ensemble $\Omega$ of graphs with maximum entropy, except that the ensemble average of the node degrees 
are constrained to the values observed for the real network: $\langle k_i \rangle_{\Omega}\equiv k_i$ $\forall i$. 
This leads to a probability distribution over $\Omega$ which is defined via a set of Lagrange multipliers $\{x_i\}$ (one for each node) 
associated to the constraints \cite{Squartini2011}. Once all $\{x_i\}$ are found, the CM reduces to having a link between nodes $i$ and $j$ with probability
$p_{ij}=\frac{x_ix_j}{1+x_ix_j}$, 
independently on all other links. Then, the null hypothesis is that any given network property $\chi$ varies in the range $\langle \chi \rangle_{\Omega}\pm\sigma_{\Omega}[\chi]$, 
where both average and standard deviation of $\chi$ over the ensemble can be obtained either analytically or numerically (by drawing sample networks from $\Omega$) \cite{Squartini2011}. 
The number of standard deviations by which the empirical and expected values of $\chi$ differ is given by the $Z$-score $Z[\chi]=(\chi-\langle \chi \rangle_{\Omega})/\sigma_{\Omega}[\chi]$: 
large positive (negative) values of $Z[\chi]$ indicate that $X$ is substantially larger (smaller) than expected, whereas, small values signal no significant deviation from the null model.

\section*{Results and Discussion}

\subsection*{Specificity, Essentiality and Codon Bias of E.coli genes}

\subsubsection*{Correlations between Codon Bias indices.} 
\begin{figure}[h!]
\includegraphics[width=0.75\textwidth]{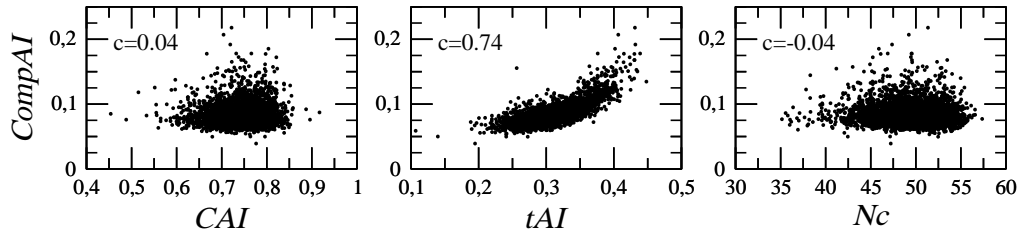}
\caption{{\bf Correlation between codon bias indices.} 
Values of Pearson's correlation coefficients show that $CompAI$ is strongly and positively correlated with $tAI$ ($c=0.74$), but not with both $CAI$ nor $Nc$ ($c\simeq 0$).}
\label{corr}
\end{figure}

As the starting point of our analysis, we first check how the different codon bias indices correlate over E.coli's genome. 
Fig. \ref{corr} shows that, interestingly, $CompAI$ is strongly (and positively) correlated with $tAI$, whereas it does not show any significant correlation with $CAI$ nor with $Nc$. 
This result can be easily explained as $CompAI$ and $tAI$ elaborate on the same genetic information, that is the abundance of tRNAs, whereas $CAI$ and $Nc$ are based on codon usage statistics 
(see the Supporting Information).

\subsubsection*{Codon Bias and ERI.} 
\begin{figure}[h!]
\includegraphics[width=0.75\textwidth]{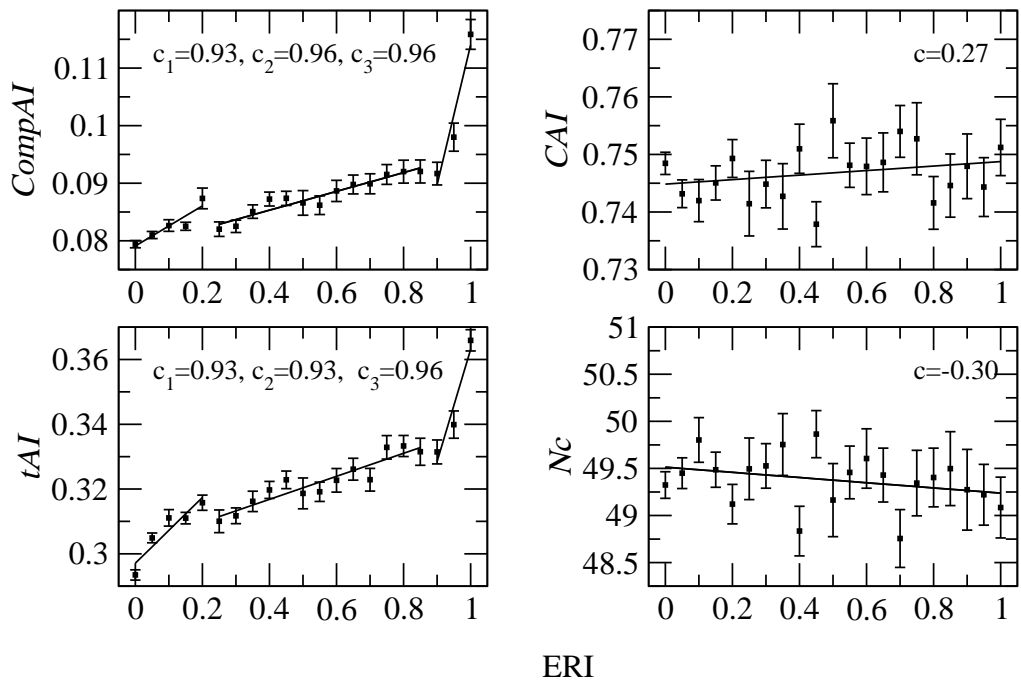}
\caption{{\bf Correlation between the various codon bias indices and ERI}. 
Codon bias average values and standard deviation (error bars) are determined for each set of E.coli genes having the same ERI value. 
In each panel, the solid lines are linear regression fits, with $c$ denoting the corresponding correlation coefficients.
In the left panels, the fits are performed separately for the three groups of genes A (ERI $<$ 0.2), B (0.2 $<$ ERI $<$ 0.9) and C (ERI $>$ 0.9). 
Both $CompAI$ and $tAI$ monotonously increase with ERI, whereas $CAI$ and $Nc$ show a low correlation with ERI.}
\label{corrERI}
\end{figure}

We move further and analyze the correlation between the various codon bias indices and the evolutionary retention index (ERI) \cite{Gerdes2003} for E.coli genes 
(we recall that a gene with a low ERI value is peculiar to E.coli, whereas a gene with high ERI is shared among different species). 
Fig. \ref{corrERI} reports the average values and standard deviations of the codon bias indices for every group of genes having the same ERI value. 
Interestingly, the evolutionary codon adaptation measured by $CompAI$ and $tAI$ tends to increase for genes that are less specific to E.coli.  
Fig. \ref{corrERI} also suggests that it is possible to make a threefold separation of genes by looking at the rate of variation of $tAI$ and $CompAI$ with ERI. 
We thus identify group A (ERI $<$ 0.2: 1597 low ERI genes that are specific to E.coli), group B (0.2 $<$ ERI $<$ 0.9: 1804 intermediate ERI genes) and group C (ERI $>$ 0.9: 231 high-ERI genes 
that are highly conserved and shared among several bacterial species). In each group, the correlation between codon bias and ERI is maximized (see the corresponding correlation coefficients in the figure). 
$CAI$ and $Nc$ are instead less structured with respect to ERI, as shown by the very small correlation coefficients (and by the impossibility to identify gene groups). 

\subsubsection*{Codon Bias and Gene Essentiality.}
\begin{figure}[h!]
\centering
\includegraphics[width=0.375\textwidth]{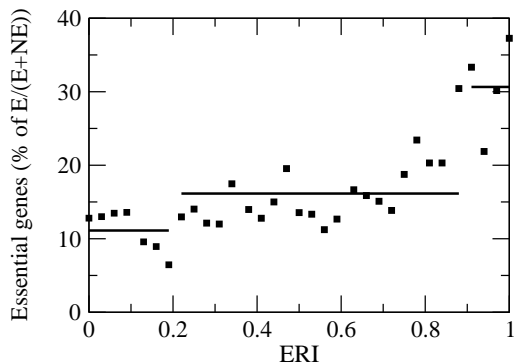}
\caption{{\bf Essentiality for E.coli genes}. The percentage of essential genes is reported for each set of genes sharing the same ERI. 
Horizontal solid lines represent average values of essentiality percentage for each group A, B, C of genes (defined by a maximum correlation between $CompAI$-ERI and $tAI$-ERI). 
The groups have different incidences of essential genes:  10\% for group A (ERI $<$ 0.2), 15\% for group B (0.2 $<$ ERI $<$ 0.9) and more than 30\% for group C (ERI $>$ 0.9).}
\label{essERI}
\end{figure}

\begin{figure}[h!]
\includegraphics[width=0.75\textwidth]{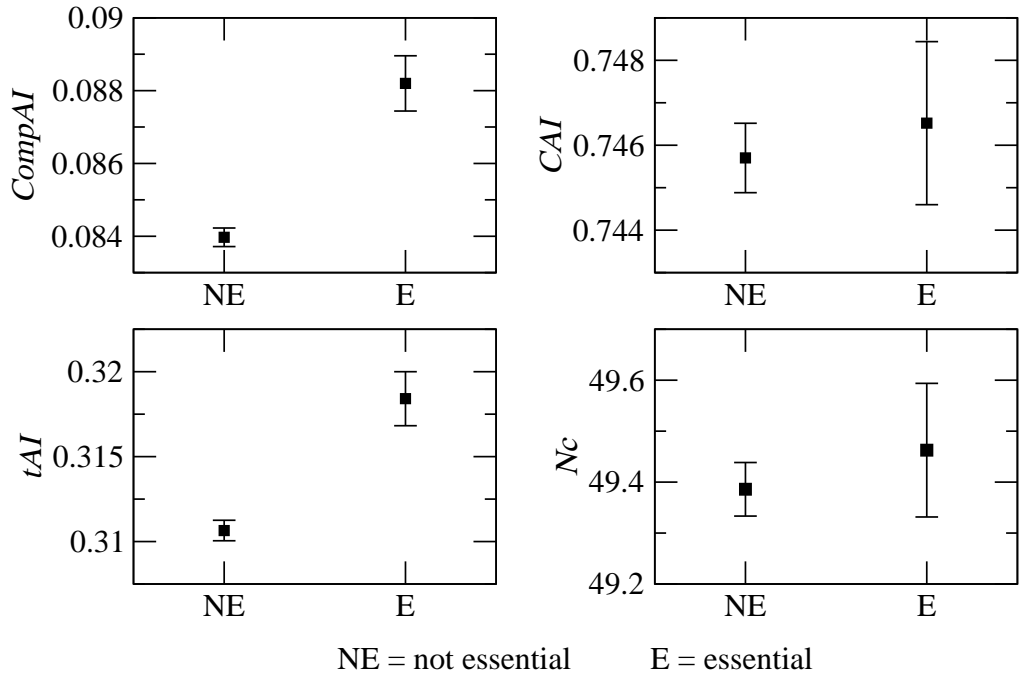}
\caption{{\bf Codon bias indices for essential and non-essential genes.} 
Error bars are standard deviations within each group. Then mean value of codon bias is systematically higher for essential genes, 
however, only $CompAI$ and $tAI$ can effectively separate essential from non-essential genes. 
In fact, in the left panels the average codon bias values for essential and non-essential genes have a relative variation of about 5\%, whereas, in the right panels 
such values are almost coincident and the errors overlap.}
\label{E-NE}
\end{figure}

We now study the patterns of codon usage bias in essential and non-essential genes, according to the classification scheme of Gerdes et al. \cite{Gerdes2003} (see Materials and Methods). 
As a preliminary result, Fig. \ref{essERI} reports the percentage of essential genes in each set of genes sharing the same ERI. 
We see that the three groups A, B, C of genes identified as in the previous paragraph feature different percentages of essential genes: approximately, 10\% for group A, 15\% for group B and above 30\% for group C. 
Essentiality and ERI thus seems to capture similar genetic features. Fig. \ref{E-NE} shows instead that $CompAI$ and $tAI$ are more sensitive 
than $CAI$ and $Nc$ in distinguishing essential from non-essential genes. Overall, Figs. \ref{corrERI} and \ref{E-NE} provide a clear indication that codon bias, as measured by $tAI$ and $CompAI$, 
is more pronounced for genes that are highly conserved ({\em i.e.}, with high ERI) and essential, on the other hand $CAI$ and $Nc$ are less sensitive to these quantities. 

\subsubsection*{COGs.} 
\begin{figure}[h!]
\includegraphics[width=0.8\textwidth]{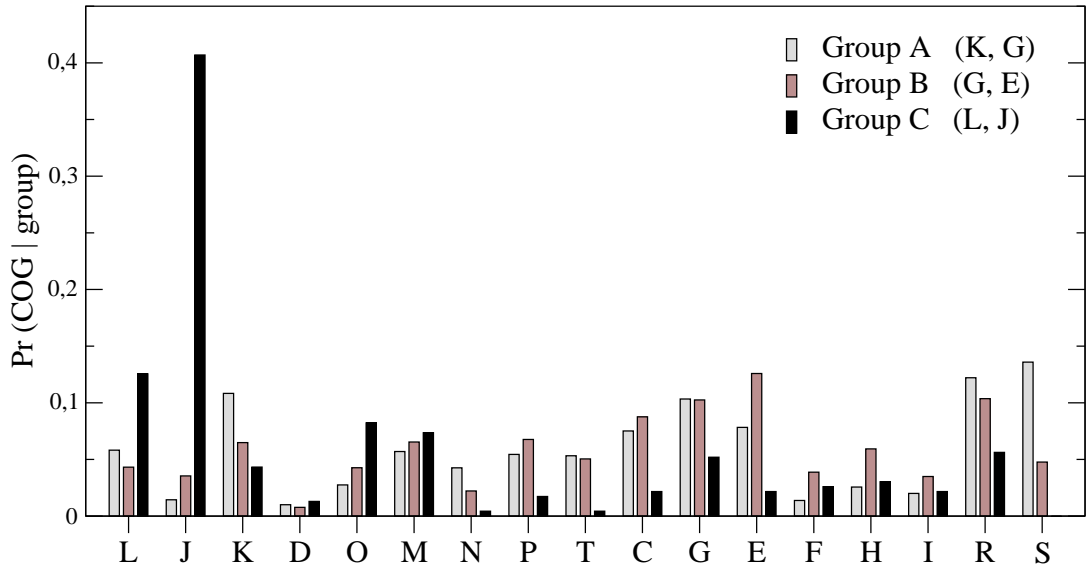}
\caption{{\bf Histogram of $ \boldsymbol{Pr(\mbox{COG}|\mbox{group})}$ over the COGs for the three gene groups A, B, C}. 
Each group is characterized by one or a few predominant COGs, indicated within parenthesis in the legend (assuming a threshold of $0.1$ and excluding generic COGs R and S, 
for which function prediction is too general or missing).}
\label{histo}
\end{figure}

We now perform a kind of gene ontology to check how the three gene groups A, B, C are projected over the clusters of orthologous genes (COGs) and their functional annotations \cite{Tatusov2001}. 
To this end, for each group we evaluate the Bayesian probability that its genes belong to a given COG: 
$Pr(\mbox{COG}|\mbox{group})=Pr(\mbox{group}|\mbox{COG})Pr(\mbox{COG})/Pr(\mbox{group})$, where $Pr(\mbox{group})$ is estimated as the fraction of the genome belonging to the group, $Pr(\mbox{COG})$ 
as the fraction of the genome belonging to the COG and $Pr(\mbox{group}|\mbox{COG})$ is the fraction of genes in the COG that belong to a particular group. 
Fig. \ref{histo} shows the histogram of $Pr(\mbox{COG}|\mbox{group})$ over the 17 COGs, for the three groups A, B, C defined above. 
Assuming an arbitrary discriminating threshold of 10\%, we observe that each group is prevalently projected over a limited set of COGs (reported in the legend box of Fig. \ref{histo}). 
Group A genes (those with low ERI values) mostly insist over COGs K and G (transcription, carbohydrate metabolism); 
group B (genes with intermediate ERI) is enriched in COGs G and E (again, carbohydrate metabolism, amino acid metabolism and transport); 
finally, group C (genes with the highest ERI) is dominated by the functional annotations associated with COGs J and L (translation, ribosome structure and biogenesis, replication, recombination and repair). 
Indeed, group C, composed of the highly adapted, essential, and conserved genes of E.coli, is the set of genes that code for ribosomal proteins.

\subsection*{Codon Bias and the Connectivity Patterns of E.coli's Protein Interaction Network}

\subsubsection*{Communities.} 
\begin{figure}[h!]
\includegraphics[width=0.75\textwidth]{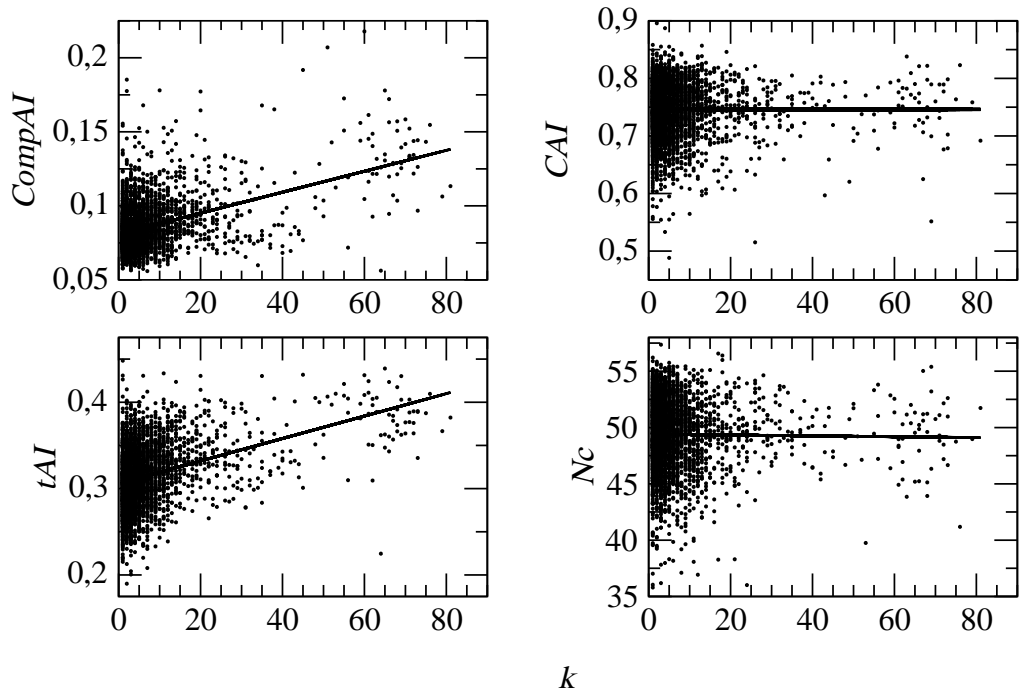}
\caption{{\bf Relation between the various codon bias indices of genes and the degree $k$ of the corresponding proteins in the PIN of E.coli}. Solid lines are linear fits. 
$CompAI$ and $tAI$ of a gene definitely increase with the connectivity of the corresponding protein in the PIN, whereas the other two indices are less sensitive to this parameter.}
\label{grado}
\end{figure}

We now turn our attention to the network of interacting proteins in E.coli. 
We start by studying codon bias in relation with the connectivity patterns of the network. First, note that the degree distribution of proteins is scale-free 
(see the right panel of Fig. \ref{fig.P_kw} of Supporting Figures), meaning that the network features a large number of poorly connected proteins and a relatively small number of highly connected hubs. 
Fig. \ref{grado} notably shows that these hub proteins are systematically characterized by higher values of codon bias of the corresponding genes---when this is measured by $tAI$ and $CompAI$. 
$CAI$ and $Nc$ are instead clearly less sensitive to protein connectivity.

We move further and consider codon bias in relation with the community structure of the PIN. We recall that a community is a group of proteins that are more densely connected within each other 
than with the rest of the network. Table \ref{tab.comm} shows the features of the communities that are assigned by MCODE a score higher than 10, 
together with their COG composition, average degree and, for ERI and the various codon bias indices, the internal average value $\bar{x}_c$ and the $Z$-scores 
(comparing the distribution of bias inside the community with that of the whole network). We see that such topologically determined communities, ordered by score, are evolutionarily and functionally characterized by a dominant COG, shared by the majority of the proteins in the community. 
This suggests that the identified communities can be associated with specific metabolic functions: they correspond to functional modules, essential for the life-cycle of the organism. 

\begin{table}[h!]
\caption{{\bf Features of top-scoring communities}. Number of nodes ($n$), community score ($n$ times the internal density), predominant COG label and percentage; 
then, for ERI and the various codon bias indices, mean values $\bar{x}_c$ internal to the community and $Z$ scores (between square brackets). Values of $Z>1$ are reported in bold.}
\begin{tabular}{|c|c|c|c|c|c|c|c|c|c|c|} 
\hline																			
ID	&	$n$	&	score &$<k>$&	COG	&	ERI	&	$CompAI$	&	$CAI$	&	$tAI$	&	$Nc$	\\
\hline\hline																			
1	&	60	&	54.9	&63.15&	J (90.0\%)	&	0.91	&	0.13	&		0.75	&	0.38	&	49.16	\\
	&		&		&&		&	[{\bf 1.66}]	&	[{\bf 1.40}]		&	[0.05]	&	[{\bf 1.35}]	&	[-0.06]	\\
\hline																			
2	&	31	&	28.6	&35.03&	N (74.2\%)	&	0.38	&	0.08		&	0.75	&	0.32	&	49.88	\\
	&		&		&&		&	[0.21]	&	[-0.35]		&	[0.07]	&	[0.14]	&	[0.12]	\\
\hline																			
3	&	21	&	19.1	&25.85&	C (97.6\%)	&	0.53	&	0.09		&	0.74	&	0.34	&	50.18	\\
	&		&		&&		&	[0.65]	&	[0.38]		&	[-0.13]	&	[0.72]	&	[0.2]	\\
\hline																			
4	&	15	&	13.9	&18.40&	M (66.7\%)	&	0.82	&	0.09		&	0.75	&	0.31	&	49.32	\\
	&		&		&&		&	[{\bf 1.31}]	&	[0.07]		&	[0.15]	&	[0.07]	&	[-0.02]	\\
\hline																			
5	&	13	&	11.7	&10.77&	P (76.9\%)	&	0.20	&	0.08		&	0.77	&	0.33	&	48.57	\\
	&		&		&&		&	[-0.29]	&	[-0.26]		&	[0.40]	&	[0.54]	&	[-0.22]	\\
\hline																			
6	&	12	&	11.5	&11.50&	U (48.9\%)	&	0.20	&	0.07		&	0.76	&	0.28	&	48.92	\\
	&		&		&&		&	[-0.29]	&	[-0.82]		&	[0.26]	&	[-0.63]	&	[-0.12]	\\
\hline																			
7	&	11	&	10.6	&19.82&	P (63.6\%)	&	0.56	&	0.09		&	0.76	&	0.34	&	48.74	\\
	&		&		&&		&	[0.70]	&	[0.44]	&	[0.26]	&	[0.72]	&	[-0.14]	\\
\hline																			
8	&	10	&	10.0	&11.60&	C (75.0\%)	&	0.04	&	0.07		&	0.76	&	0.29	&	47.78	\\
	&		&		&&		&	[-0.86]	&	[-0.66]	&	[0.28]	&	[-0.45]	&	[-0.30]	\\
\hline
\end{tabular}
\label{tab.comm}
\end{table}

Let us focus on the first community, that includes only 60 proteins (4.5\% of the whole network) but as much as 32.6\% of the total number of links in the network, 
and that basically overlaps with the main core of the PIN ({\em i.e.}, the $k$-core with the highest possible degree). Notably, proteins belonging to this community have on average a codon bias index 
(as measured by $tAI$ and, even more, by $CompAI$) that is significantly higher than the average of the rest of the network (the $Z$-score is bigger than 1). 
As noticed above, this core is essentially composed of ribosomal proteins, that are usually highly expressed, have the highest codon usage bias, 
and are broadly conserved and essential across different taxa \cite{Butland2005}. 

\subsubsection*{Principal Component Analysis.} 
\begin{figure}[h!]
\includegraphics[width=0.5\textwidth]{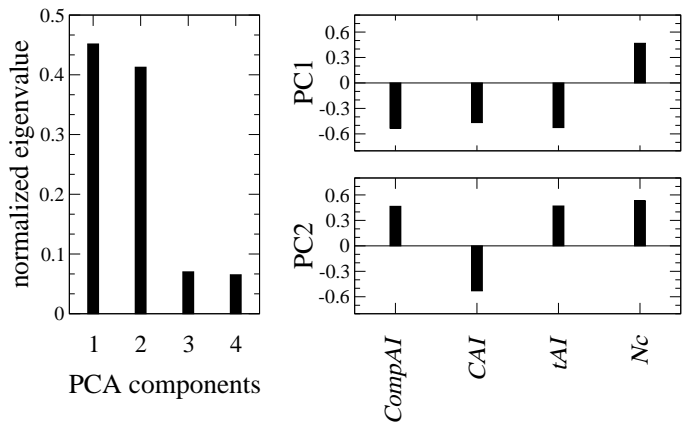}
\caption{Left plot: {\bf Eigenvalues of the correlation matrix between the codon bias indices on expressed sequences.} 
Right plot: {\bf Projection of the first two PCA components on the individual codon bias indices.} Recalling that $Nc$ is anticorrelated with the other codon bias indices, 
PC1 results from a weighted and coherent contribution of all the indices, whereas, for PC2 the contribution of $CompAI$ and $tAI$ is opposite to that of $CAI$ and $Nc$.}
\label{PCA}
\end{figure}

\begin{figure}[h!]
\includegraphics[width=0.375\textwidth]{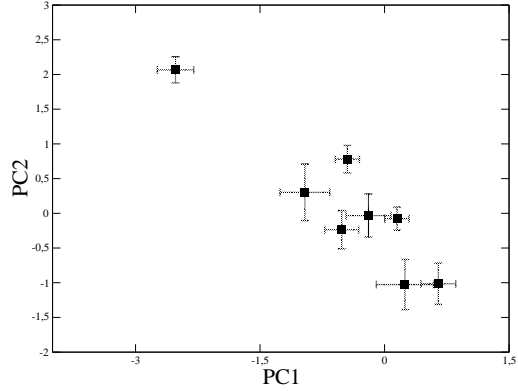}
\caption{{\bf Centroids of the top MCODE communities in the space of the first two PCA components.} The error bars denote the variance of the centroids.}\label{PCA_comm}
\end{figure}

\begin{figure}[h!]
\includegraphics[width=0.5\textwidth]{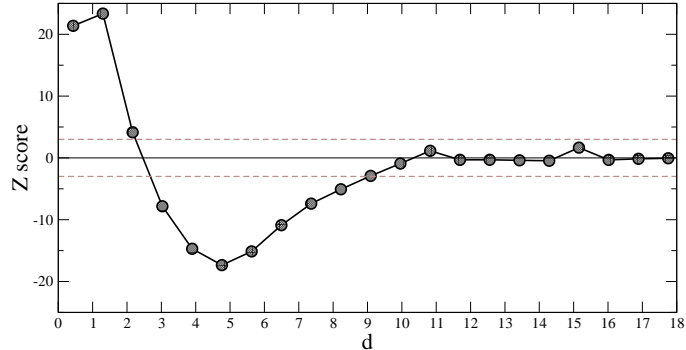}
\caption{{\bf Histogram of the $\boldsymbol Z$-score for $ \boldsymbol{Pr(\mbox{link}|d)}$ for each pair of genes and their respectively encoded proteins.} 
$d$ is the Euclidean distance between pairs of genes in the space of the first two PCA components, and $Pr(\mbox{link}|d)$ is the conditional probability of having a link in the PIN between two proteins 
given that their encoding genes are localized within a distances $d$ in the PC1-PC2 plane. The $Z$-score is obtained as 
$Z[Pr(\mbox{link}|d)]=[Pr(\mbox{link}|d)-\langle Pr(\mbox{link}|d) \rangle_{\Omega}]/\sigma_{\Omega}[Pr(\mbox{link}|d)]$. 
The gray dashed lines mark the significance interval of $\pm3\sigma$.}
\label{P_link}
\end{figure}

Finally we perform PCA  over the space of the four codon bias indices ($CompAI$, $CAI$, $tAI$, $Nc$) 
measured for each E.coli gene. The two first principal components (PC1 and PC2) turn out to represent for as much as 85\% of the total variance of codon bias over the genome (left plot of Fig. \ref{PCA}). 
Projection of the first two principal components on the individual codon bias indices (loadings) shows that none of the four indices predominantly contributes to the data variability (right plot of Fig. \ref{PCA}). 
Thus, the placement of a gene in the PC1-PC2 plane depends on a weighted contribution of all the indices. Interestingly, the genes encoding for the proteins of the eight top MCODE communities 
are well localized and separated in this reduced space (Fig. \ref{PCA_comm}). In particular, the first community ({\em i.e.}, the core of ribosomal proteins characterized by high values of both $CompAI$ and $tAI$) 
is located in the upper left part of the graph, isolated from the others. This represents an important evidence: proteins that belong to the densest connected cores of the interactome 
are well-localized in the space of the two principal components. 
In other words, if a set of proteins are physically and functionally connected in a module, then their corresponding genes should share common codon bias features. 
Conversely, we can obtain an estimate for the conditional probability $Pr(\mbox{link}|d)$ of a functional interaction between proteins, provided that their relative genes fall within a distance $d$ 
in the plane of the two principal components PC1 and PC2. Reasonably, we compare $Pr(\mbox{link}|d)$ estimated on the real interactome with $\langle Pr(link|d)\rangle_\Omega$ estimated 
on the Configuration Model (CM) which, we recall, is a degree-conserving  randomization (re-wiring) of the network. 
Fig. \ref{P_link} shows the $Z$-score for $Pr(\mbox{link}|d$) as a function of $d$, and reveals a peculiar behavior: for small distances ($d\le2$) the probability of finding a connection between 
two proteins is much higher than what could have been expected from a (degree-conserving) random link placement. Conversely, for medium distances ($3\le d\le9$), 
the linking probability is lower than that of the CM, whereas, the real network and the CM become compatible for large distances, where, however, connections are rather few. 
This analysis shows that sets of genes sharing similar codon usage patterns encode for proteins that are much more likely to interact than in situations where chance alone is responsible 
for the structure of the interactome. 

\section*{Conclusion}

In this work we have introduced $CompAI$, a novel codon bias index that is inspired by $tAI$, though conceptually distinct. 
In fact, $CompAI$ does not make reference to lists of highly expressed genes, and is thus unsupervised and based on intrinsic information about co-evolution of genes that code for proteins and tRNAs. 
Conceptually, the definition of $CompAI$ is based on a model that postulates a competition between cognate and near-cognate tRNAs for the same codon, exposed on the ribosome at each step of protein synthesis. 
Competitive mechanisms in the machinery of ribosomal translation of genes into proteins have been repeatedly suggested and studied in the literature \cite{Fluitt2007} and deserve further attention 
in order to understand their role for translation efficiency. In particular, $CompAI$ was designed in order to provide information about the speed of protein synthesis, being based on proofreading delay mechanisms.

Our genome-wide analysis of codon bias in E.coli using $CompAI$ as well as other commonly used indices revealed that codon usage metrics resting on counting tRNA genes ($CompAI$ and $tAI$) 
are strongly and positively correlated among themselves---in spite of their conceptually different definition. 
It would then be quite interesting to check in the future whether this correlation is specific to E.coli or it is universally observed in the genomes of bacterial species 
that are either ecologically and evolutionarily close or, by contrast, very far from E.coli. 
We also found that both $CompAI$ and $tAI$ correlate with ERI, the degree of conservation for a gene among similar species, and gene essentiality, whereas, $CAI$ and $Nc$ are less sensitive to these quantities. 
$CompAI$ and $tAI$ values also allow to distinguishing three groups of genes, that are differently characterized by both codon choice adaptation, ERI and degree of essentiality, 
and that also feature specific predominant COG signatures. In particular the third group (C), composed of the few genes that are highly conserved and with the strongest codon bias adaptation, 
consists for 30\% of essential genes with predominant COGs J and L---that refer to translation, ribosome structure and biogenesis, replication, recombination and repair. 
These represent house-keeping and control functions that must be continuously executed by the cell, meaning that the genes responsible for them have to be expressed most of the time during the cell cycle. 
These observations strongly support the idea that an increasing selection of codons and, in parallel, 
a correlated modulation of tRNA availability co-evolved along the evolutionary history of a species. 

Finally, we have addressed a theme as relevant as the connection between codon usage bias and protein functional or physical interactions. 
Our main result indicates that, in the course of the evolution of a genome, the functional structuring of the complex of interactions between proteins has interfered 
with the peculiar codon-coding formulation of the corresponding genes. In particular we have shown here, for the first time to our knowledge, that communities of highly connected proteins 
in the interactome of E.coli correspond to encoding genes that share the same degree of evolutionary adaptation, as expressed by codon bias indices 
that synthetically represent genetic information encoded in the tRNAs sector of the genome. Indeed, $CompAI$, that is based on a simple representation of tRNA competition, 
seems to detect the codon bias signal behind communities more consistently than the other indices here considered. Conversely, we have provided evidence that if two genes have similar codon usage patterns 
then the corresponding proteins have a significant probability of being functionally connected and interacting. This result points out that codon bias should be a relevant parameter in the fundamental problem of predicting unknown protein-protein interactions from genomic information. 
This study is a first exploratory step towards a more complete investigation on how communities within protein-protein interaction networks rest 
on a consistent but still to be decoded codon bias signal.  Indeed, the connection of the topology of a network with an underneath semantics is far from trivial, as recently pointed out in the specialized literature \cite{Hric2014}. Biological PINs and codon bias offer an interesting case study worth to be investigated in a wider perspective \cite{Boccaletti2014}.


\setcounter{figure}{0}
\setcounter{section}{0}
\setcounter{equation}{0}

\makeatletter 
\renewcommand{\thefigure}{S\@arabic\c@figure}
\renewcommand{\thesection}{S\@arabic\c@section}
\renewcommand{\theequation}{S\@arabic\c@equation}
\makeatother

\section*{Supporting Information}
\label{Supporting Information}

Here we define and explain the rationale behind the various codon bias indices that have been proposed in the literature.
\newline

{\bf Codon Adaptation Index} ($CAI$) \cite{Sharp1987}. The pattern of codon usage is ruled by two simultaneous effects \cite{Bulmer1991}: 
translational selection towards optimal codons for each amino acid, and genetic drift that allows the persistence of non-optimal codons. 
It is natural to assume that selection is stronger for codons of highly expressed genes, which thus feature a more pronounced bias in the use of codons. 
The principle behind $CAI$ is exactly that codon usage in highly expressed genes can reveal the optimal ({\em i.e.}, most efficient for translation) codons for each amino acid. 
Hence, $CAI$ builds on a reference set of highly expressed genes to assess, for each codon $i$, the relative synonymous codon usages ($RSCU_i$) and the relative codon adaptiveness ($w_i$):
\begin{equation}
RSCU_i=\frac{X_i}{\frac{1}{n_i}\overset{n_i}{\underset{j=1}{\sum}}X_j};\qquad\qquad w_i=\frac{RSCU_i}{\displaystyle\max_{j=1,\dots,n_i}\,\{RSCU_j\}};\label{eq:RSCU_w_cai}
\end{equation}
In the $RSCU_i$, $X_i$ is the number of occurrences of codon $i$ in the genome, and the sum in the denominator runs over the $n_i$ synonyms of $i$; 
$RSCU$s thus measure codon usage bias within a family of synonymous codons.
$w_i$ is then defined as the usage frequency of codon $i$ compared to that of the optimal codon for the same amino acid encoded by $i$---{\em i.e.}, the one which is mostly used 
in a reference set of highly expressed genes. 

The $CAI$ for a given gene $g$ is calculated as the geometric mean of the usage frequencies of codons in that gene, 
normalized to the maximum $CAI$ value possible for a gene with the same composition of amino acid:
\begin{equation}
CAI_g=\left(\overset{l_g}{\underset{i=1}{\prod}}w_i\right)^{1/l_g},\label{eq:cai}
\end{equation}
where the product runs over the $l_g$ codons belonging to that gene (except the stop codon). 
The critical aspect in the definition of $CAI$ is that it requires to define {\em a priori} reference set of highly expressed genes that is specific for a given organism. 
$CAI$ is then not always transferable; yet, since it is tuned on highly expressed genes, it is generally well correlated with gene expression levels in genomes for which reference gene sets are available. 
In this work, $CAI$ for E.coli genes was computed using the DAMBE 5.0 package \cite{Xia2013}
\newline

{\bf tRNA Adaptation Index} ($tAI$) \cite{dosReis2004}. The speed of protein synthesis is bound to the waiting time for the correct tRNA to enter the ribosomal A site \cite{Varenne1984}, 
and thus depends on tRNA concentrations \cite{Soerensen1989} (and, indirectly, on gene copy numbers). The consequent adaptation of codon usage to tRNA availability \cite{Ikemura1981,Ikemura1985} 
is at the basis of $tAI$, which follows the same mathematical model of $CAI$---defining for each codon $i$ its absolute ($W_i$) and relative ($w_i$) adaptiveness value:
\begin{equation}
W_i=\overset{m_i}{\underset{j=1}{\sum}}\left(1-s_ {ij}\right)\mbox{tGCN}_{ij};\qquad\qquad w_ i=\begin{cases}
W_ i/W_ {max}&\mbox{if $W_i\neq0$}\\
w_{mean}&\mbox{otherwise}
\end{cases};\label{eq:W_w_tAI}
\end {equation}
where $m_i$ is the number of isoacceptor tRNAs that recognize codon $i$ ({\em i.e.}, tRNAs that carry the same aminoacid that is encoded by $i$), 
tGCN$_{ij}$ is the gene copy number of the $j$-th tRNA that recognizes the $i$-th codon, 
$s_{ij}$ is a selective constraint on the efficiency of the codon-anticodon coupling, $W_{max}$ is the maximum $W_i$ value and $w_{mean}$ is the geometric mean of all $w_i$ with $W_i\neq0$. 

The $tAI$ of gene $g$ is eventually defined as the geometric mean of the relative adaptiveness values of its codons, thus estimating the amount of adaptation of gene $g$ to its genomic tRNA pool:
\begin{equation}
tAI_g=\left(\overset{l_g}{\underset{i=1}{\prod}}w_i\right)^{1/l_g}.\label{eq:tAI}
\end{equation}
The critical issue for $tAI$ is the selection of a meaningful set of $s_{ij}$ values, \emph{i.e.}, weights that represent wobble interactions between codons and tRNAs. 
Assuming that tRNA usage is maximal for highly expressed genes, these values are chosen in order to optimize the correlation of $tAI$ values with expression levels---exactly as $CAI$. 
Besides, while the efficiencies of the different codon-tRNA interactions are expected to vary among different organisms, 
$s_{ij}$ values are based on the gene expression in {\em Saccharomyces cerevisiae} \cite{dosReis2004}---thus lacking universality \cite{Sabi2014}.
In this work we have evaluated $tAI$ values of E.coli genes using the CodonR package \cite{CodonR}.
\newline

{\bf Effective Number of Codons} ($Nc$) \cite{Wright1990}. $Nc$ is a measure that quantifies the departure of a gene from the random usage of synonymous codons. 
Given a sequence of interest, the computation of $Nc$ \cite{Sun2012} starts from the quantity---defined for each family $\alpha$ of synonymous codons:
\begin{equation}
F_{CF_\alpha}=\overset{m_\alpha}{\underset{k=1}{\sum}}\left(\frac{n_{k_\alpha}}{n_\alpha}\right)^2\label{eq:F_cf}
\end{equation}
where $m_\alpha$ is the number of codons in $\alpha$ (each appearing $n_{1_\alpha},n_{2_\alpha},\dots,n_{m_\alpha}$ times in the sequence) and $n_\alpha=\sum^{m_\alpha}_{k=1}n_{k_\alpha}$. 
$Nc$ then weights these quantities in order to measure amount of entropy in the codon usage of the sequence:
\begin{equation}
Nc=N_S+\frac{K_2\,\overset{K_2}{\underset{\alpha=1}{\sum}}n_\alpha}{\overset{K_2}{\underset{\alpha=1}{\sum}}\left(n_\alpha\,F_{CF_\alpha}\right)}+\frac{K_3\,\overset{K_3}{\underset{\alpha=1}{\sum}}n_\alpha}{\overset{K_3}{\underset{\alpha=1}{\sum}}\left(n_\alpha\,F_{CF_\alpha}\right)}+\frac{K_4\,\overset{K_4}{\underset{\alpha=1}{\sum}}n_\alpha}{\overset{K_4}{\underset{\alpha=1}{\sum}}\left(n_\alpha\,F_{CF_\alpha}\right)}\label{eq:N_c}
\end{equation}
where $N_S$ is the number of families with one codon only and $K_m$ is the number of families with degeneration $m$ (families with degeneration 6 are divided into two families 
of degeneration 2 and 4, as they often are subject to different selective forces). Note that $Nc$ reaches is maximal value (61) 
when all codons are used equally and its minimal value (23) when only one codon is used per amino acid (extreme bias). 
Differently from both $CAI$ and $tAI$, $Nc$ is a more immediate measure of codon usage that does not require any {\em a priori} information nor makes any biological hypothesis 
(which constitute its weakness and, and the same time, its strength). Yet, since the effect of selection is a reduction of entropy for codon usage in a sequence, $Nc$ provides a reliable measure for this effect. 
In this paper we have obtained $Nc$ values through DAMBE 5.0 \cite{Xia2013}.

\section*{Supporting Figures}

\begin{figure}[h!]
\includegraphics[width=1.0\textwidth]{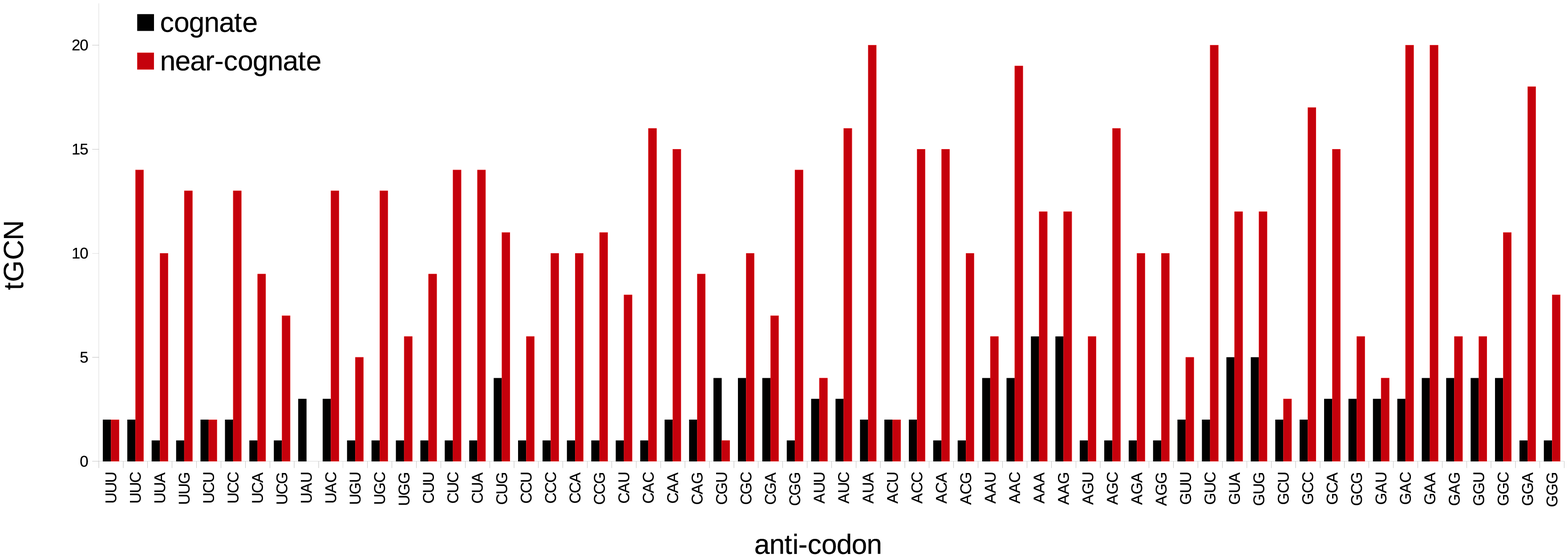}
\caption{{\bf Abundance of tGCN cognate and near cognate (according to Watson-Crick base pairing) for each anti-codon in E.coli.} Data taken from \cite{Chan2009}}\label{fig.cognate}
\end{figure}

\begin{figure}[h!]
\includegraphics[width=1.0\textwidth]{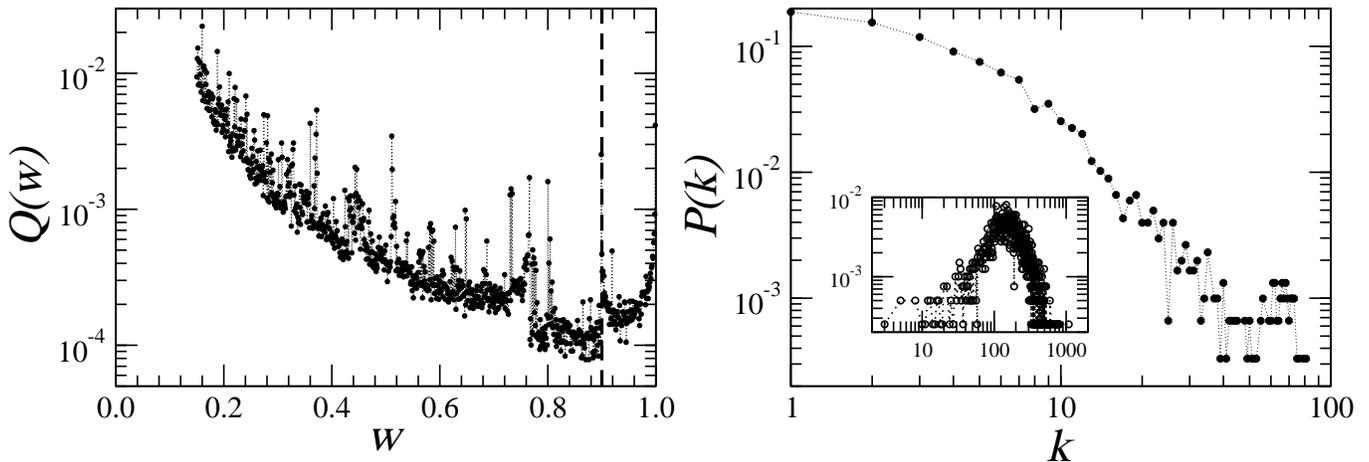}
\caption{Left plot: {\bf Distribution of confidence levels} $Q(w)$, with the vertical line indicating the cut-off we use to separate true from false positives. 
Right plot: {\bf Distribution of degrees} $P(k)$ when $\Theta=0.9$, with the insets showing the same distribution for the original network ($\Theta=0$).}\label{fig.P_kw}
\end{figure}




\begin{thebibliography}{}
\bibitem{foot1} The density of a graph $G$ with $n$ nodes and $l$ links is the ratio between $l$ and the maximum number of possible links, namely $n(n-1)/2$, 
whereas, a $k$-core is a graph $G$ of minimal degree $k$, meaning that each node belonging to $G$ has degree greater or equal than $k$.
\bibitem{foot3} Here is the functional classification of COGs: 
L, replication, recombination and repair; J, translation, including ribosome structure and biogenesis; K, transcription;
 D, cell division and chromosome partitioning;
 O, molecular chaperons and related functions;
 M, cell wall structure, biogenesis and outer membrane;
 N, secretion, motility and chemotaxis;
 P, inorganic ion transport and metabolism;
 T, signal transduction;
 C, energy production and conversion;
 G,carbohydrate metabolism and transport;
 E, amino acid metabolism and transport; 
 F, nucleotide metabolism and transport; 
 H, coenzyme metabolism;
 I, lipid metabolism;
 R, general function prediction only;
 S, no function prediction.
\end{thebibliography}
\end{document}